\documentclass[10pt,prd,twocolumn,floatfix,preprintnumbers,nofootinbib,superscriptaddress,longbibliography]{revtex4-1}
\pdfoutput=1
\usepackage[colorlinks=true,breaklinks=true]{hyperref}
\hypersetup{allcolors=[rgb]{0.0 0.0 0.6},linkcolor=[rgb]{0.0 0.5 0.6}}
\usepackage[normalem]{ulem}
\usepackage{amsmath,amssymb}
\usepackage{epsfig}  
\usepackage{graphicx}   
\usepackage{slashed}             
\usepackage{url}
\usepackage{color}
\usepackage{multirow}
\usepackage{placeins}
\usepackage[dvipsnames]{xcolor}
\usepackage{letltxmacro}
\LetLtxMacro{\oldcite}{\cite}
\renewcommand{\cite}[1]{\mbox{\oldcite{#1}}}
\newcommand{\gag}{g_{a\gamma}}
\usepackage{comment}
\usepackage{bm}

\DeclareMathOperator{\muG}{\mu G}
\DeclareMathOperator{\GeV}{GeV}
\DeclareMathOperator{\eV}{eV}

\DeclareMathOperator{\im}{Im}
\DeclareMathOperator{\re}{Re}

\DeclareMathOperator{\kpc}{kpc}

\DeclareMathOperator{\Log}{Log}

\newcommand{\bk}{{\bf k}}

\newcommand{\bB}{{\bf B}}

\newcommand{\bn}{\hat{n}}

\newcommand{\beq}{\begin{equation}}
\newcommand{\eeq}{\end{equation}}

\newcommand{\uu}{\bm{u}}
\newcommand{\ff}{\bm{f}}
\newcommand{\BB}{\bm{B}}
\newcommand{\JJ}{\bm{J}}
\newcommand{\AAA}{\bm{A}}
\newcommand{\DD}{{\rm D} {}}
\newcommand{\SSSS}{\mbox{\boldmath ${\sf S}$} {}}
\newcommand{\nab}{{\bm{\nabla}}}

\def\Rm{\mbox{\rm Re}_{\rm M}}
\def\Rey{\mbox{\rm Re}}
\def\Lu{\mbox{\rm Lu}}
\def\Ma{\mbox{\rm Ma}}
\def\urms{u_{\rm rms}}
\def\Brms{B_{\rm rms}}
\def\kf{k_{\rm f}}
\def\cs{c_{\rm s}}
\def\vA{v_{\rm A}}
\def\vvA{\bm{v}_{\rm A}}
\def\kurt{\mbox{\rm kurt}}
\newcommand{\bra}[1]{\langle #1\rangle}

\usepackage{colortbl}
\definecolor{rp}{cmyk}{0.2, 1, 0.6, 0}
\definecolor{green2}{cmyk}{0.25, 0.74, 0.86,0}

\allowdisplaybreaks

\bibpunct{[}{]}{,}{n}{}{,}

\begin{document}

\title{Magnetohydrodynamics predicts heavy-tailed distributions of axion-photon conversion}

\author{Pierluca Carenza}
\email{pierluca.carenza@fysik.su.se}
\affiliation{The Oskar Klein Centre, Department of Physics, Stockholm University, 10691 Stockholm, Sweden}
\author{Ramkishor Sharma}
\email{ramkishor.sharma@su.se}
\affiliation{Nordita, KTH Royal Institute of Technology and Stockholm University, Hannes Alfv\'ens v\"ag 12, 10691 Stockholm, Sweden}
\affiliation{The Oskar Klein Centre, Department of Astronomy, Stockholm University, 10691 Stockholm, Sweden}
\author{M.C.~David Marsh}
\email{david.marsh@fysik.su.se}
\affiliation{The Oskar Klein Centre, Department of Physics, Stockholm University, 10691 Stockholm, Sweden}
\author{Axel Brandenburg}
\email{brandenb@nordita.org}
\affiliation{Nordita, KTH Royal Institute of Technology and Stockholm University, Hannes Alfv\'ens v\"ag 12, 10691 Stockholm, Sweden}
\affiliation{The Oskar Klein Centre, Department of Astronomy, Stockholm University, 10691 Stockholm, Sweden}
\affiliation{McWilliams Center for Cosmology and Department of Physics, Carnegie Mellon University, 5000 Forbes Ave, Pittsburgh, PA 15213, USA}
\affiliation{School of Natural Sciences and Medicine, Ilia State University, 3-5 Cholokashvili Avenue, 0194 Tbilisi, Georgia}
\author{Eike Ravensburg}
\email{eike.muller@fysik.su.se}
\affiliation{The Oskar Klein Centre, Department of Physics, Stockholm University, 10691 Stockholm, Sweden}

\date{\today}
\smallskip

\begin{abstract}
The interconversion of axionlike particles (ALPs) and photons in magnetised astrophysical environments provides a promising route to search for ALPs. The strongest limits to date on light ALPs use galaxy clusters as ALP--photon converters. However, such studies traditionally rely on simple models of the cluster magnetic fields, with the state-of-the-art being Gaussian random fields (GRFs).  We present the first systematic study of ALP--photon conversion in more realistic, turbulent fields from dedicated magnetohydrodynamic (MHD) simulations, which we compare with GRF models. For GRFs, we analytically derive the distribution of conversion ratios at fixed energy and find that it follows an exponential law. We find that the MHD models agree with the exponential law for typical, small-amplitude mixings but exhibit distinctly heavy tails for rare and large mixings. We explain how non-Gaussian features, e.g.~coherent structures and local spikes in the MHD magnetic field, are  responsible for the heavy tail.
Our results suggest that limits placed on ALPs using GRFs are robust.
\end{abstract}

\maketitle

\section{Introduction}
Axion-like particles (ALPs) are ubiquitous in extensions of the Standard Model of particle physics~\cite{Peccei:1977hh, Wilczek:1977pj, Svrcek:2006yi, Jaeckel:2010ni,Ringwald:2014vqa,DiLuzio:2020wdo} and provides an increasingly popular candidate for dark matter \cite{Preskill:1982cy, Abbott:1982af, Dine:1982ah}. ALPs and photons can interconvert in background magnetic fields through the interaction \cite{Sikivie:1983ip, Raffelt:1987im} 
\begin{equation}
    \mathcal{L}=-\frac{g_{a\gamma}}{4}a\tilde{F}^{\mu\nu}F_{\mu\nu}\,,
    \label{eq:lagr}
\end{equation}
where $a$ is the ALP field, $\gag$ is the ALP--photon coupling, $F_{\mu\nu}$ is the electromagnetic tensor, and $\tilde{F}^{\mu\nu}=\frac{1}{2}\epsilon^{\mu\nu\rho\sigma}F_{\rho\sigma}$ is its dual. Laboratory searches based on this interaction include light-shining-through-the-wall experiments, `helioscopes' that are sensitive to solar ALPs, and dark matter `haloscopes', see \cite{Graham:2015ouw,Sikivie:2020zpn,Workman:2022ynf} for reviews. Moreover, complementary astrophysical searches for ALP--photon mixing have proven exceptionally powerful and have led to some of the strongest limits on $\gag$, as we now discuss. 

The strength of the ALP--photon mixing grows with the magnitude of the magnetic field and the size of the magnetised region. Galaxy clusters are both magnetised (with $\mu$G field strengths) and large (spanning hundreds of kiloparsecs), and are known to be efficient ALP--photon converters: a significant fraction of high-energy photons travelling through a cluster could emerge as ALPs, and vice versa. It follows that the high-energy photon spectra of a bright Active Galactic Nucleus (AGN) located within or behind a galaxy cluster would be distorted by ALP--photon conversions, which makes it possible to use X-ray and gamma-ray satellites to search for ALPs. For example, precision X-ray spectrometry of the cluster-hosted quasar H1821+643 limits the amplitude of possible ALP-induced distortions to the $\lesssim 2.5\%$ level, which implies strong constraints on ALP theories \cite{Reynes:2021bpe}. A comparable precision has been reached for the AGN NGC1275 at the centre of the Perseus cluster \cite{Reynolds:2019uqt}, and the several similar studies have been performed in the X-ray and gamma-ray ranges  \cite{Hooper:2007bq,  horns2012, Wouters:2013hua, Conlon:2013txa, Berg:2016ese,Marsh:2017yvc,Chen:2017mjf, Conlon:2017qcw, Conlon:2017ofb,   HESS:2013udx, Cicoli:2014bfa, Meyer:2014epa,Meyer:2014gta, Fermi-LAT:2016nkz,  Zhang:2018wpc,Malyshev:2018rsh,  Galanti:2018upl,  Conlon:2018iwn,  Bu:2019qqg,Day:2019ucy, Guo:2020kiq, Cheng:2020bhr,Carenza:2021alz, Matthews:2022gqi, Kachelriess:2021rzc, Kachelriess:2021rzc, Schallmoser:2021sba, Galanti:2022yxn, Galanti:2022tow,  Jacobsen:2022swa, Buen-Abad:2020zbd,Li:2020pcn,Li:2021gxs,Li:2022jgi}

A critical step in translating the absence of spectral distortions into limits on ALPs is the modelling of the cluster magnetic field. For some clusters, Faraday rotation measure studies constrain key properties of the magnetic field, but the detailed spectral shape of the ALP--photon conversion ratio is given by the full spatial autocorrelation function of the magnetic field~\cite{Marsh:2021ajy}, which is not observationally accessible. The magnetic field must therefore be treated as a \emph{nuisance parameter} to be marginalised over in the statistical analysis. Traditionally, rather simple models for the cluster magnetic field have been used in the literature, e.g.~by taking the field to be constant within a series of cells along the direction of propagation (i.e.~`cell-models'), or by using divergence-free Gaussian random fields (GRFs). Recent studies have found the predictions to be rather robust to the choice of model \cite{Matthews:2022gqi}. Still, it is important to note that these simple magnetic field models differ significantly from more realistic models found by solving the dynamical magnetohydrodynamic (MHD) equations of motion. In particular, MHD simulations of turbulent plasmas exhibit \emph{coherent magnetic structures} on moderately large scales, which have no counterpart for GRF or cell-model fields. This prompts the question: can the predictions of the simpler models be trusted, given their lack of coherent structure?

This paper presents the first systematic study of ALP--photon conversion in MHD models of cluster magnetic fields (see also Ref.~\cite{Montanino:2017ara} for a similar study with ENZO code).
We carefully analyse the predicted distributions of ALP--photon conversion ratios and compare them to those of similar GRF models. Importantly, we find that the \emph{typical} predictions are in excellent agreement between the MHD and GRF models, but that the distributions differ significantly for \emph{rare} fluctuations. While the fluctuations to large conversion ratios are exponentially suppressed in GRF models, the MHD models predict distinctly heavy tails. We establish the origin of the heavy tails to be coherent structures with large amplitude magnetic fields, which reflect the non-Gaussianity of the MHD field. By contrast, we find the helicity of the magnetic field to be unimportant for the conversion ratio. 

In Sec.~\ref{sec:oscillations} we summarise the formalism of ALP-photon conversions in the perturbative regime. In Sec.~\ref{sec:MHDmodels} we discuss how MHD simulations model the environment encountered in galaxy clusters. In the literature, these turbulent magnetic fields are modeled with simplified GRF models, discussed in Sec.~\ref{sec:GRFmodels}. In Sec.~\ref{sec:results} we compare the results obtained for ALP-photon conversions in these different magentic field models and conclude.

\section{ALP--photon oscillations}
\label{sec:oscillations}
The linearised, classical field equations of electromagnetism and relativistic ALPs is given by a Schr\"odinger-like equation~\cite{Raffelt:1987im}:
\begin{equation}
i \frac{d}{dz} \Psi(z) = H\Psi(z)\, ,
\label{eq:EoM0}
\end{equation}
where $z$ denotes the spatial coordinate along the direction of propagation. The components of the three-level `state vector' $\Psi(z)=(A_x, A_y, a)^{T}$ consists of the transverse components of the vector potential, i.e.~`photon', and the ALP field. In the presence of a background magnetic field, ${\bf B}({\bf x})$, the ALP--photon interaction of Eq.~\eqref{eq:lagr} induces off-diagonal elements that mix the ALP with the photons, 
\begin{equation}
    H=\begin{pmatrix}
\Delta_\gamma& 0 & \Delta_{a\gamma_x}\\
0 & \Delta_\gamma & \Delta_{a\gamma_y} \\
\Delta_{a\gamma_x} & \Delta_{a\gamma_y} & \Delta_a
\end{pmatrix}\,, 
\label{eq:mat}
\end{equation}
where $\Delta_{a\gamma_i}=g_{a\gamma} B_{i}/2$ for $i=x,\, y$, $\Delta_{\gamma}=-\frac{\omega_{\rm pl}^{2}}{2\omega}$ and $\Delta_{a}=-\frac{m_{a}^{2}}{2\omega}$. We have neglected Faraday rotation and the QED birefringence effect, which are both negligible at X-ray energies. 

Due to the mixing by $\Delta_{a\gamma_i}$, a  photon beam of energy $\omega$ that is linearly polarised along the $i$-direction loses a fraction of $P_{a\gamma_i}(\omega)$ of its flux into ALPs, where $P_{a\gamma_i}$ denotes the `quantum mechanical' conversion probability calculated from Eq.~\eqref{eq:EoM0}. Throughout this paper, we refer to $P_{a\gamma}$ as the conversion ratio. In the bulk of this paper, we focus on the most instructive, `massive' regime where $m_a\gg\omega_{\rm pl}$, but 
the general conclusions hold for arbitrary ALP masses (see App.~\ref{App:massless}). This is not surprising, given that the mixing of ALPs and photons in an environment with constant plasma frequency is formally identical to the massive regime, and that the mixing for varying plasma frequencies results in well-understood, and typically mild, modifications to the calculations.

Throughout this paper, we will be interested in the small mixing regime where this problem can be solved perturbatively in $g_{a\gamma}$~\cite{Raffelt:1987im}, resulting in the  leading order transition amplitude 
\begin{equation}
   i {\cal A}_{\gamma_{i}\to a} = \frac{g_{a\gamma}}{2} \int_{-L/2}^{L/2} dz\,
    B_i(z\hat z) e^{-i\eta_{a}z} \equiv \frac{g_{a\gamma}}{2} \tilde B_i(\eta_a) \,,
    \label{eq:fourier}
\end{equation}
where 
\begin{equation}
\eta_{a}=-\Delta_{a} = 0.078\, \text{kpc}^{-1}\, \left(\frac{m_a}{10^{-12}\, \text{eV}}\right)^2 \left(\frac{\text{keV}}{\omega} \right)
\end{equation}
and where, without loss of generality, the trajectory goes through the origin of the coordinate system. We discuss the conversion ratio for unpolarised fluxes in App.~\ref{App:helicity}.
Clearly, as $L \to \infty$, the perturbative amplitude is the one-dimensional Fourier transform of the relevant component of the magnetic field along the $z$-direction, evaluated at the conjugate `momentum' $\eta_a$. It follows that the conversion ratio is given by the one-dimensional power spectrum of the magnetic field, 
\begin{equation}
    P_{a\gamma_i}(\eta_a) = \frac{g_{a\gamma}^2}{4} |\tilde B_i(\eta_a)|^2 \, ,
    \label{eq:Ppower}
\end{equation}
and can also be expressed as the Fourier transform of the real-space magnetic autocorrelation function \cite{Marsh:2021ajy}. The Fourier representation makes it possible to efficiently evaluate the conversion ratio using celebrated numerical methods such as the Fast Fourier Transform \cite{Marsh:2021ajy, Matthews:2022gqi, Reynes:2021bpe}. Moreover, Eq.~\eqref{eq:Ppower} immediately suggests a subtle conceptual question. Extended magnetic structures rely on phase coherence, and disappear when the phases of the Fourier components are randomised.\footnote{See, e.g.~Figs.~18--21 of \cite{Maron:2000gh}.}  The conversion ratio is only a function of the norm of $\tilde B_i$, so \emph{is $P_{a\gamma_i}$ independent of coherent structures?} The answer to this question is no, as we will now explicitly demonstrate by comparing the predictions of magnetic field models from three-dimensional MHD simulations to GRF models.

\begin{figure*}[t]
\vspace{0.cm}
\includegraphics[width=\textwidth]{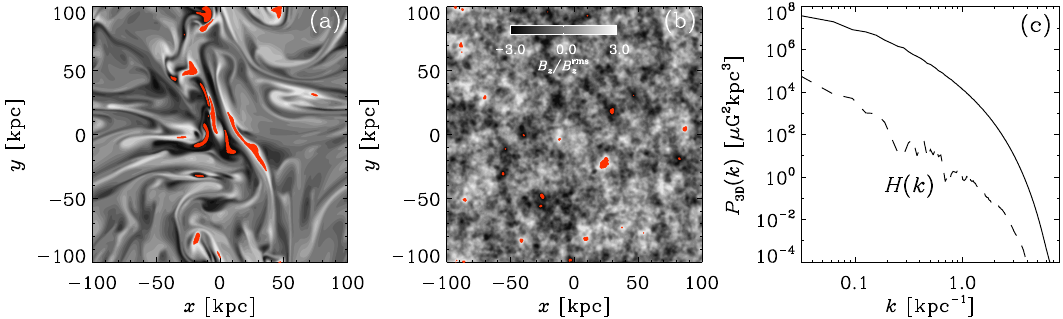}
\caption{
Cross-section of the magnetic field for (a) Run $\mathcal{S}$ and (b) a GRF model with same the power spectrum. The red regions highlight locations with $|B|>3\Brms$. (c) Power and helicity spectra for Run $\mathcal{S}$. 
}\label{fig:Hfig}
\end{figure*}

\section{MHD models of the magnetic field}
\label{sec:MHDmodels}
The intra-cluster medium (ICM) is a dilute, magnetised plasma with tangled structures on few-kiloparsec scales, as is evident from radio and X-ray observations. The ICM is near local hydrostatic equilibrium and is characterised by a very high magnetic Reynolds number ($\sim 10^{30}$) and significant MHD turbulence.  The magnetic field is thought to be generated through gas motions by a dynamo process, driven by a variety of internal and external processes such as magneto-thermal instabilities \cite{Scheko+05, balbus:2010, Parrish+12}, jet-activity from the central AGN resulting in ICM shocks and cavities \cite{Branduardi-Raymont:1981}, turbulent wakes of individual galaxies \cite{Ruzmaikin+89,Subramanian+06, ruszkowski:2010}, and sporadic cluster mergers~\cite{Xu+09,Vazza+12}.

To analyse ALP--photon conversion in dynamically generated magnetic fields, we perform state-of-the-art MHD simulations using the {\sc Pencil Code}~\cite{JOSS} in a box of size $L^3= (200\, \text{kpc})^3$ with $512^3$ mesh points. The turbulence is assumed to be driven through some external volume forcing, which we model as random sinusoidal waves that are $\delta$-correlated in time, i.e.~the forcing function changes at each time step. The wave vectors are taken from a shell of finite thickness and radius $\kf$, which we chose to be close to the smallest wave number of the computational domain $k_1\equiv2\pi/L$.  This results in turbulence at a moderate magnetic Reynolds number that is as large as possible for the given numerical resolution. The ratio of viscosity to magnetic diffusivity is 20. We ignore the density stratification and just consider an isothermal gas with constant sound speed. The simulated magnetic fields do not decrease with radius, as is expected in galaxy clusters, and should not be interpreted as fully realistic models. Rather, the MHD models exhibit turbulence and structure, as are expected in real galaxy clusters, and allow us to test the robustness of the ALP theory predictions.
  
We initialise the simulations with a weak seed magnetic field. After about 50 turnover times ($\urms\kf t=10$ where $\urms$ is the rms velocity), the magnetic field begins to grow exponentially. During this phase, the magnetic field is highly non-Gaussian, but the field strength is still weak. To assess the consequences of such a highly non-Gaussian field, we consider a scaled version of this magnetic field, referred to as Run $\mathcal{K}$, because the dynamo is kinematic, i.e.~unaffected by magnetic feedback.

When the magnetic energy density reaches values comparable to the kinetic energy density, the Lorentz force begins to affect the turbulence and leads to a saturation of the dynamo. We refer to this state as Run $\mathcal{S}$. The magnetic field is then still non-Gaussian, but the kurtosis is smaller than during the kinematic stage. The density also becomes more strongly affected by the magnetic field. In both runs, the turbulence is isotropic on large length scales to a good approximation (see App.~\ref{App:MHD}).

To study the properties of ALP--photon conversion in the MHD magnetic fields, we consider the ensemble of trajectories defined by straight lines in the $z$-direction through the simulation volume. The corresponding ensemble of conversion ratios is then defined by solving Eq.~\eqref{eq:EoM0} along each trajectory for a range of energies. In practice, this procedure is drastically simplified by the perturbative formalism, as the relevant predictions can be efficiently extracted from the one-dimensional power spectrum of the magnetic field, cf.\ Eq.~\eqref{eq:Ppower}. This way, Runs~$\mathcal{K}$ and $\mathcal{S}$ result in two distinct statistical distributions of conversion ratios, which we now compare with the predictions of GRF models.  

\section{GRF models of the magnetic field}
\label{sec:GRFmodels}
GRFs provide a convenient mathematical framework for tailoring smooth magnetic fields with an arbitrary power spectrum. GRF models have been used to study the properties of cluster magnetic fields based on Faraday rotations \cite{Ensslin:2003ez, Bonafede:2010}, and have also been used in ALP searches in the X-ray and gamma-ray energy ranges \cite{Horns:2012kw, Angus:2013sua, Meyer:2014epa, Fermi-LAT:2016nkz,Matthews:2022gqi, Reynes:2021bpe, Schallmoser:2021sba}. In this section, we analytically compute the distribution of ALP--photon conversion ratios in GRF magnetic field models. 

We consider photons that are linearly polarised along the $i$-direction and propagate along an ensemble of rays in the $z$-direction. The perturbative conversion ratio is given by Eq.~\eqref{eq:Ppower}, and our goal in this section is to determine the statistical distribution, $f_{P(\eta_a)}$, of $P_{a\gamma_i}$ at fixed $\eta_a$.

	\begin{figure*}[t!]
		\vspace{0.cm}
		\includegraphics[width=0.45\linewidth]{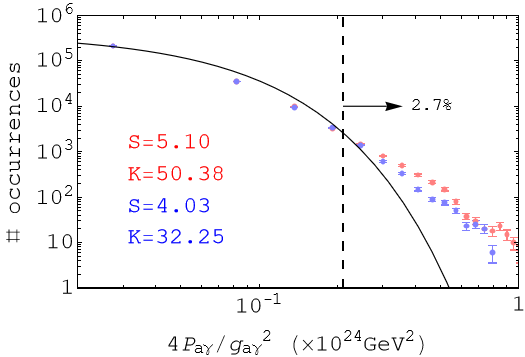}
		\includegraphics[width=0.45\linewidth]{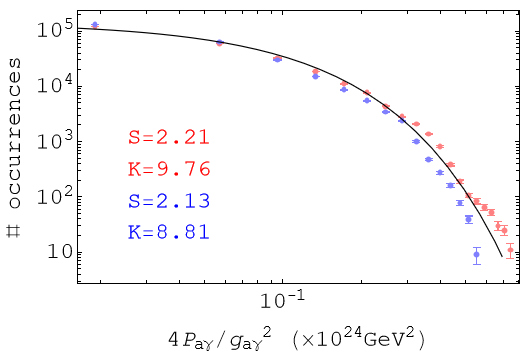}			\includegraphics[width=0.45\linewidth]{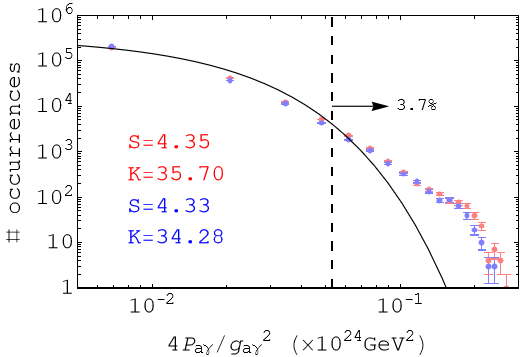}
		\includegraphics[width=0.45\linewidth]{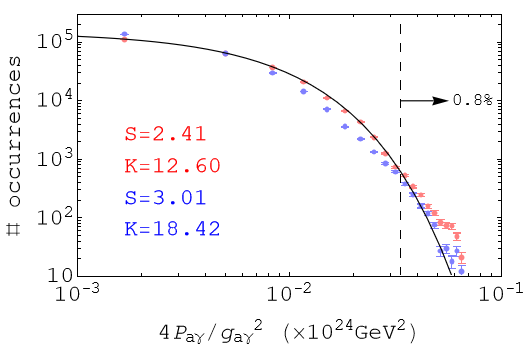}
			\vspace{0.cm}
		\includegraphics[width=0.45\linewidth]{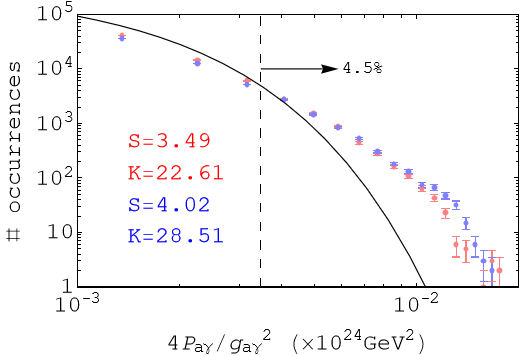}
		\includegraphics[width=0.45\linewidth]{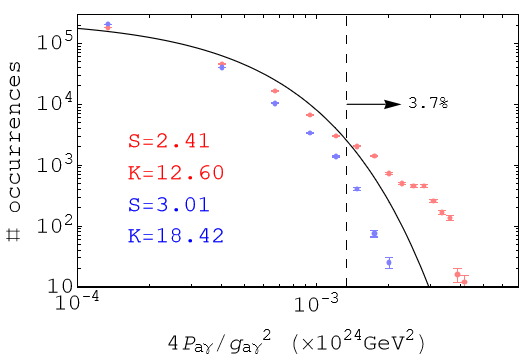}
\caption{Histogram of the conversion ratio extracted from Runs~$\mathcal{K}$ (left column) and $\mathcal{S}$  (right column) for an initial photon flux that is linearly polarised along the $x$ (blue) and $y$ (red) directions together with the analytical GRF prediction of Eqs.~\eqref{eq:1D3D}--\eqref{eq:exp} (black line). Here $\eta_{a}=7.8\times10^{-3}~\kpc^{-1}$ (top row), $\eta_{a}=0.07~\kpc^{-1}$ (middle row), $\eta_{a}=0.195~\kpc^{-1}$ (bottom row). Skewness $S$ and kurtosis $K$ of the distributions are shown with the corresponding colour. For comparison, $S=2$ and $K=9$ for an exponential distribution.   Approximate locations and cumulative probabilities of the heavy tails are respectively indicated by dashed vertical lines and percentages. The error bars are obtained by assuming Poissonian fluctuations.} \label{fig:probdMHD}
\end{figure*}

In order to isolate the intrinsic differences between models from MHD and GRF, we compare magnetic field models that have the same three-dimensional power spectrum. We denote the three-dimensional Fourier transform of the magnetic field as
\begin{equation}
    \widehat B_a({\bf k}) = \int d^3 {\bf x}\, B_a({\bf x})\, e^{-i {\bf k } \cdot {\bf x}} \, ,
\end{equation}
where the index $a$ runs over all three spatial coordinates. For a magnetic field that is statistically homogeneous and isotropic, as our MHD and GRF magnetic field models, the two-point correlation function is given by
\cite{Ensslin:2003ez}
\begin{equation}
\begin{split}
    \langle \widehat B_{a}(\bk) \widehat B_{b}^{*}(\bk')\rangle&=\delta^{3}(\bk-\bk')\\
    &\left[M_{N}(k)\left(\delta_{ab}-\frac{k_{a}k_{b}}{k^{2}}\right)-i\epsilon_{abc}\frac{k_{c}}{k}H(k)\right]\,,
    \label{eq:2point}
\end{split}
\end{equation}
where $M_N$ and $H$ respectively are the normal and helical autocorrelation functions, following the conventions of \cite{Subramanian:1999mr}. The three-dimensional power spectrum is given by the trace of the autocorrelation tensor: $P_{\rm 3D}(k) = 2   M_{N}(k)$. The autocorrelation of $\tilde B_i$ is then given by
\begin{equation}
\langle \tilde B_i(\eta_a) \tilde B^*_j(\eta_a')\rangle
= \delta(\eta_a-\eta_a')\Big[ \delta_{ij} P_{\rm 1D}(\eta_a) + i \epsilon_{ijz} H_{\rm 1D}(\eta_a) \Big] \,,
\label{eq:tildeBcorr}
\end{equation}
where 
   \begin{equation}
   \begin{split}
 &P_{\rm 1D}(\eta_{a})=\int \frac{dk_{\perp}k_{\perp}}{2(2\pi)^{3}}P_{\rm 3D}\left(\sqrt{\eta_{a}^2+k_{\perp}^2}\right)\left(1-\frac{1}{2}\frac{k_{\perp}^{2}}{\eta_{a}^{2}+k_{\perp}^{2}}\right)\,,
   \\
   & H_{\rm 1D}(\eta_a) = \int \frac{dk_{\perp}k_{\perp}}{(2\pi)^{3}}H\left(\sqrt{\eta_{a}^2+k_{\perp}^2}\right)\frac{\eta_a}{\sqrt{\eta_{a}^{2}+k_{\perp}^{2}}} \, , \label{eq:1D3D} 
\end{split} 
\end{equation}
where $k_\perp = \sqrt{k_x^2 + k_y^2}$.  The delta-function prefactor in Eq.~\eqref{eq:tildeBcorr} arises in the $L\to \infty$ limit of the expressions $\sin\big[L(\eta_a- k) \big]/(2\pi(\eta_a-k)) \approx \delta(\eta_a-k)$ for $k= k_z, \eta'_a$. For finite but large $L$ and $\eta=\eta'$, the factor $\delta(\eta_a - \eta_a')$ is regulated to $L/(2\pi)$. Clearly, helicity is unimportant for the conversion ratio of linearly polarised photons (see App.~\ref{App:helicity}).
		
The one-dimensional power spectrum $P_{\rm 1D}$ fully determines the statistical properties of the Gaussian magnetic field, and thus also those of the conversion ratio. Following the derivation in, e.g.~Ref.~\cite{Carenza:2021alz}, we determine the probability density function (PDF), $f_{P(\eta_a)}$, of the random variable $P_{a\gamma}(\eta_{a})$ in the ensemble defined by the Gaussian magnetic fields (see App.~\ref{App:GRF} for GRFs and App.~\ref{App:nonG} for the non-Gaussian case).
The resulting PDF takes a very simple, exponential form:
\begin{equation}
\begin{split}
    f_{P_{a\gamma}(\eta_{a})}(p)&=\frac{e^{-p/p_{0}}}{p_{0}},\quad
    p_{0}=\frac{g_{a\gamma}^{2}}{4}\frac{L}{2\pi}P_{\rm 1D}(\eta_{a})\, .
    \label{eq:exp}
\end{split}
\end{equation}
This equation provides an ideal starting point to compare the predictions of GRFs to those of MHD simulations: we numerically extract the three-dimensional magnetic power spectrum from the MHD runs, and use this in Eqs.~\eqref{eq:1D3D} and  \eqref{eq:exp} to determine the semi-analytical GRF prediction for $f_{P_{a\gamma}}$ at fixed $\eta_a$. We then compare this prediction with the empirical distribution of conversion ratios calculated from Eq.~\eqref{eq:Ppower} for the ensemble of trajectories through the MHD runs.

\section{Results and conclusions}
\label{sec:results}
Figure~\ref{fig:Hfig} shows a cross-section of the magnetic field for Run $\mathcal{S}$ in the left panel, the extracted power spectrum ($P_{\rm 3D}$) and the helicity ($H$) in the right panel, and a realisation of a GRF magnetic field with the same $P_{\rm 3D}$ in the central panel.  

Clearly, fields are visually very different due to the appearance of coherent structures in the MHD simulation. For each run we used $512^{2}$ straight lines as photon paths to calculate the conversion ratio using the perturbative formalism, cf.\ Eq.~\eqref{eq:Ppower}, which we checked to be in agreement with a complete numerical solution (see Ref.~\cite{Calore:2020tjw, Matthews:2022gqi}) at the $\sim 0.1\%$ level in the relevant regime.\footnote{Nearby trajectories are separated by a distance smaller than the coherence length of the magnetic field and cannot be regarded as statistically independent. However, we have checked that the predictions agree well with less densely sampled trajectories, and do not expect this to significantly affect our results. In addition, note that the chosen trajectories cross the entire simulation box (of length $200$~kpc), equispaced and sampled in the innermost $190$~kpc of a box face to avoid boundary effects.}

Each row of Fig.~\ref{fig:probdMHD} shows, for a fixed $\eta_a$, the distribution (PDF) of $P_{a\gamma}(\eta_{a})$ in Run $\mathcal{K}$ (left column) and Run $\mathcal{S}$ (right column) together with the analytical, GRF prediction from Eq.~\eqref{eq:exp} (black solid line). The red and blue points correspond to different linear polarisations of the photon flux (i.e.~$i=x,\, y$). The agreement between the red and blue histograms for typical, small values of the probability is a reflection of the approximate statistical isotropy of the MHD magnetic field. 

All models agree well for the prediction of the small-amplitude conversion ratios that are the most common, and the MHD runs reproduce the exponential decay of the GRF field.
However, small-amplitude features in observational photon spectra can be buried by systematic uncertainties and measurement errors.
For sufficiently large conversion ratios, the predictions from MHD runs differ significantly from the GRF predictions: for rare fluctuations, the MHD distributions exhibit a spectral break followed by a distinctly heavy tail. Consequently, large conversion ratios are more frequent in MHD models than in GRFs. We note that the average conversion probability depends only on $P_{\rm 3D}$ and is, therefore, the same for MHD and GRF models. This explains the  additional, fractionally-small difference, visible in Fig.~\ref{fig:probdMHD}, between the GRF and MHD conversion ratios at intermediate values of $4P_{a\gamma}/g_{a\gamma}^2$. We note that the appearance of the heavy tails does not originate from poor statistics, as is evident from the small error bars of the data points in Fig.~\ref{fig:probdMHD}.

Unsurprisingly, the highly non-Gaussian magnetic field from the kinematic phase of Run $\mathcal{K}$ results in the largest deviations from the GRF predictions. In this case, a strong deviation from the exponential law occurs in $1$--$5\%$ of the realisations for each of the sampled $\eta_a$. The distribution from the saturated phase of Run $\mathcal{S}$ deviates from the GRF predictions more strongly at low $\eta_a$ (i.e.~high energy). 

The independence of $P_{a\gamma}$ on the phases of $\tilde B$, cf.~Eq.~\eqref{eq:Ppower}, ensures that the mean value, $\langle P_{a\gamma}\rangle$, is fixed by the power spectrum and is therefore the same for the MHD and GRF models. However, the heavy tails are reflected in enhanced higher moments of $P_{a\gamma}$. 

Quantitatively for a random variable $\mu$ with mean $\bar{\mu}$ and variance $\sigma^{2}$, we indicate the the skewness as $S=\langle (\mu-\bar{\mu})^{3}\rangle/\sigma^{3}$ and the kurtosis as $K=\langle (\mu-\bar{\mu})^{4}\rangle/\sigma^{4}$. These quantities are shown in each panel of Fig.~\ref{fig:Hfig} corresponding to the conversion ratio PDFs. In most cases, both $S$ and $K$ are significantly larger than the values for an exponential distribution of the GRFs, $S=2$ and $K=9$.

There are two distinct non-Gaussian features that can contribute to the physical origin of the heavy tails of $f_{P(\eta_a)}$: the presence of extended coherent structures, and larger amplitude peaks (spikes) of the magnetic field compared to a GRF. In the context of TeV-scale ALP--photon conversion in magnetic fields from cosmological simulations, Ref.~\cite{Montanino:2017ara} suggested that coherent structures (alone) can explain heavy-tailed distributions. We show that both effects contribute, see App.~\ref{App:whyfat}. However, in our MHD simulations, peaks of the magnetic field tend to be located in coherent structures, and we expect the effects to be correlated.

In many cases, the energy-dependent conversion ratio for a \emph{single} sightline is of direct observational interest. It goes beyond the scope of this paper to discuss the effect of MHD magnetic fields on the limits placed on ALPs as this would require significantly extended  modelling (e.g.~of the, on average, radially decreasing plasma frequency and magnetic field strength in a galaxy cluster) and data analysis (using real data, e.g.~from precision X-ray observations, cf.~\cite{Reynolds:2019uqt}). However, in Appendix \ref{app:statist}, we construct sets of mock X-ray observations and ask: what is the probability that a random line-of-sight taken from an MHD model is `sufficiently different' from an analogous GRF realisation? In more detail,  we consider a hypothetical point-like source in a galaxy cluster and compare the energy-dependent residuals obtained from MHD and GRF models after ALP-photon interconversion. The (mock) observed photon spectra are generated by multiplying a power-law spectrum, representing the primary X-ray flux, by photon survival probabilities calculated, respectively, with the MHD or GRF models. The mock spectra are fitted by a power-law and the statistical distribution of the residuals from the GRF models is determined. We then use the Anderson-Darling test statistic~\cite{doi:10.1080/01621459.1987.10478517} to quantify the differences between the MHD and GRF models. For $5$--$8\%$ of the MHD spectra, we can reject the null hypothesis that the residuals come from the GRF distribution at 95\% confidence level; for the vast majority of realisations, the Anderson-Darling statistic does not suffice to distinguish MHD models from the GRF distribution.  This conclusion holds even for experiments with high energy resolution ($\sim\mathcal{O}(\eV)$), like \emph{Chandra} diffraction grating observations or the future \emph{Athena} mission~\cite{Nandra:2013jka, Conlon:2017ofb, Sisk-Reynes:2022sqd}. It may be possible to improve the sensitivity to MHD effects by designing dedicated statistical tests, however, we expect GRF modelling to suffice for most applications.
%

In conclusion, we have presented the first systematic study of ALP--photon conversions in magnetic fields obtained by turbulent MHD simulations and compared the predictions to those of simpler GRF models with the same power spectrum. We showed that the typical predictions agree between these models, even though the non-Gaussian fields from MHD simulations result in heavy-tailed distributions of conversion ratios, corresponding to larger-than-expected oscillations in photon spectra from sources embedded in galaxy clusters. 
This effect is typically small for observables sensitive only to single sightlines through a cluster, in which case the GRF modelling is expected to be sufficient.

An important future direction is to study MHD models of realistic cluster magnetic fields (including the radial decrease of the field strength), and to test the robustness of published limits. Moreover, it would be interesting to further develop our cluster MHD models by including, e.g.~anisotropic viscosity \cite{Berlok+20} and apply these to searches for ALPs in observational X-ray data.

\vspace{2mm}
\subsection*{Data availability.} 
The source code used for the
simulations of this study, the {\sc Pencil Code},
is freely available from Ref.~\cite{JOSS}.
The simulation setups and the corresponding data along with animations for Runs~${\mathcal K}$ and ${\mathcal S}$
are freely available from Ref.~\cite{DATA}.

\vspace{2mm}
\acknowledgements
We would like to thank Riccardo Z.~Ferreira for discussions. 
The work of PC, DM and EM is supported by the European Research Council under Grant No.~742104 and by the Swedish Research Council (VR) under grants  2018-03641 and 2019-02337. 
The work of RS and AB is supported by VR grant 2019-04234.
Nordita is sponsored by Nordforsk.
We acknowledge the allocation of computing resources provided by the
Swedish National Allocations Committee at the Center for Parallel
Computers at the Royal Institute of Technology in Stockholm and
Link\"oping.

\onecolumngrid
\appendix
\newpage

\section{Gaussian random field (GRF)}
\label{App:GRF}
In this appendix, we briefly review how GRFs with a given power spectrum can be explicitly generated, and how the probability distribution of the conversion ratio is derived. Since the conversion ratio is determined by the one-dimensional Fourier transform, we consider a one-dimensional lattice of $2N+1$ points with spacing $dz=L/(2N+1)$ along the direction of propagation.  At each lattice point, we generate independent, random numbers, $W_{h}$, from a normal distribution with mean zero and unit variance
\begin{equation}
    f_{W_{h}}(w)=\frac{e^{-\frac{w^{2}}{2}}}{\sqrt{2\pi}}\,.
\end{equation}
The discrete Fourier transform of this white noise is given by
\begin{equation}
    \tilde{W}_{j}=\frac{1}{\sqrt{2N+1}}\sum_{h=-N}^{N}W_{h}e^{\frac{2\pi i}{N}jh}\,.
    \label{eq:Wtilde}
\end{equation}
To construct a GRF with a given one-dimensional power spectrum, $P_{\rm 1D}(\eta)$, we define the Fourier transform of this field as
\begin{equation}
    \tilde{B}_{j}=\sqrt{\tilde{P}_{\rm 1D}(\eta_{j})}\tilde{W}_{j}\,,
    \label{eq:gaussB}
\end{equation}
where $\eta_{j}=2\pi |j|/(N\,dz)$ and $\tilde{P}_{\rm 1D}(\eta_{j})=(L/2\pi)P_{\rm 1D}(\eta_{j})$.
The components in Eq.~\eqref{eq:gaussB} directly determine the ALP--photon conversion ratio 
\begin{equation}
    P_{a\gamma}(\eta_{j})=\frac{g_{a\gamma}^{2}}{4}\left|\tilde{B}_{j}\right|^{2}\,,
\end{equation}
and the statistical distribution of $P_{a\gamma}(\eta_{j})$ (at fixed $\eta_j$) only depends on the distribution of $\tilde B_j$ (at fixed $j$). The moments of the conversion ratio are given by
\begin{equation}
    \langle P_{a\gamma}(\eta_{j})^{n} \rangle=n!\left(\frac{g_{a\gamma}}{2}\right)^{2n}\tilde{P}_{\rm 1D}(\eta_{j})^{n}\,.
\end{equation}
Given the moments, the characteristic function (the expectation value of  $e^{it P_{a\gamma}(\eta_{j})}$) is easy to evaluate,
\begin{equation}
    \langle e^{it P_{a\gamma}(\eta_{j})}\rangle  
    =\int_0^\infty dp\, f_{P_{a\gamma}(\eta_{j})}(p)e^{itp}
    =\sum_{n=0}^{\infty}\frac{(it)^{n}}{n!}\langle \tilde{P}_{a\gamma}(\eta_{j})^{n}\rangle=\frac{1}{1-it \left(\frac{g_{a\gamma}}{2}\right)^{2}\tilde{P}_{\rm 1D}(\eta_{j})}\, ,
\end{equation}
and the probability distribution of the perturbative  conversion ratio can be extracted by a simple Fourier transform,
\begin{equation}
\begin{split}
    f_{P_{a\gamma}(\eta_{a})}(p)&=\frac{e^{-p/p_{0}}}{p_{0}},\quad
    p_{0}=\frac{g_{a\gamma}^{2}}{4}\tilde{P}_{\rm 1D}(\eta_{a})\, .
    \label{eq:probdistr0}
\end{split}
\end{equation}

\section{Non-Gaussian random field}
\label{App:nonG}
In this appendix,  we construct an analytically tractable example of a non-Gaussian magnetic field which we then use to demonstrate how  non-Gaussianity leads to heavy tails for the ALP--photon conversion ratio. Non-Gaussian fields can be constructed in many ways. Here, we follow the approach of Ref.~\cite{Vio:2001cm} and construct a non-Gaussian random field through a non-linear operation on a GRF. We denote the Gaussian field by $\tilde{W}_{j}$, defined as in Eq.~\eqref{eq:Wtilde}, and take the non-Gaussian field to be given by
\begin{equation}
    \tilde{B}_{j}=\sqrt{\tilde{P}_{\rm 1D}(\eta_{j})}f_{\epsilon}(\tilde{W}_{j})\,,
\end{equation}
where, for $j=1,... N$,
\begin{equation}
   f_{\epsilon}(\tilde{W}_j)=\frac{1}{1+\frac{\epsilon^{2}}{2}}\Big( \tilde{W}_j+\epsilon\, \im(\tilde{W}_j) \Big)\,.
       \label{eq:nonG}
\end{equation}
The $j=0$ mode is defined as $f_{\epsilon}(\tilde W_0) = \tilde W_0$ and the remaining, negative modes are fixed by the reality condition for the magnetic field in real space $\tilde{B}_{j}^{*}=\tilde{B}_{-j}$. We now consider $j>0$ unless explicitly stated otherwise. The factor of $\sqrt{1+\frac{\epsilon^{2}}{2}}$ in Eq.~\eqref{eq:nonG} ensures that the average magnetic field does not depend on the parameter controlling the strength of the non-Gaussianity,  $\epsilon$. The real and imaginary parts of the non-Gaussian field are explicitly given by
\begin{equation}
\begin{split}
    \re(\tilde{B}_{j})&=\frac{\sqrt{\tilde{P}_{\rm 1D}(\eta_{j})}}{\sqrt{1+\frac{\epsilon^{2}}{2}}}\left(\re(\tilde{W}_{j})+\epsilon\, \im(\tilde{W}_{j})\right)\,,\\
    \im(\tilde{B}_{j})&=\frac{\sqrt{\tilde{P}_{\rm 1D}(\eta_{j})}}{\sqrt{1+\frac{\epsilon^{2}}{2}}}\im(\tilde{W}_{j})\,.
    \end{split}
\end{equation}
It is possible to derive the PDF of the non-Gaussian field $\tilde{B}_{j}$ for an arbitrary power spectrum of $\tilde W$. In general, if $n$ random variables $\xi_{i}$ follow a distribution $f_{\bf \xi}(\xi_{1}, ... \xi_{n})$, then the new variables defined as $\psi_{j}=g_{j}(\xi_{1}, ... , \xi_{n})$ have the following PDF
\begin{equation}
    f_{\bf \psi}(\psi_{1}, \ldots , \psi_{n})=f_{{\bf \xi}}(\xi_{1}, \ldots , \xi_{n})\left|{\rm det}\frac{\partial g}{\partial \xi}\right|_{\xi_{i}=g^{-1}(\psi_{1}, \ldots, \psi_{n})}^{-1}\, .
\end{equation}
From Eq.~\eqref{eq:nonG}, the inverse transformation is
\begin{equation}
    \begin{split}
       \re(\tilde{W}_{j})&=\sqrt{1+\frac{\epsilon^{2}}{2}}\frac{\re(\tilde{B}_{j})-\epsilon \im(\tilde{B}_{j})}{\sqrt{\tilde{P}_{\rm 1D}(\eta_{j})}}\,,\\
      \im(\tilde{W}_{j})&=\sqrt{1+\frac{\epsilon^{2}}{2}}\frac{\im(\tilde{B}_{j})}{\sqrt{\tilde{P}_{\rm 1D}(\eta_{j})}}\,,\\
    \end{split}
\end{equation}
and the inverse Jacobian is
\begin{equation}
\begin{split}
    J^{-1}&=\frac{1+\frac{\epsilon^{2}}{2}}{\tilde{P}_{\rm 1D}(\eta_{j})}\,,
    \end{split}
\end{equation}
then the PDF of $\tilde{B}_{j}$ is
\begin{equation}
\begin{split}
    f_{\tilde{B}_{j}}&=\frac{1+\frac{\epsilon^{2}}{2}}{\pi \tilde{P}_{\rm 1D}(\eta_{j})} \, 
    {\rm exp} \Bigg[
    -\left(1+\frac{\epsilon^{2}}{2}\right)\frac{\left(\re(\tilde{B}_{j})-\epsilon \im(\tilde{B}_{j})\right)^{2} + \im(\tilde{B}_{j})^{2}}{\tilde{P}_{\rm 1D}(\eta_{j})} 
    \Bigg]
    \,.
    \end{split}
    \label{eq:distr}
\end{equation}
Clearly, the Gaussian distribution is recovered as $\epsilon \to 0$.
		\begin{figure}[t!]
		\vspace{0.cm}
		\includegraphics[width=0.45\linewidth]{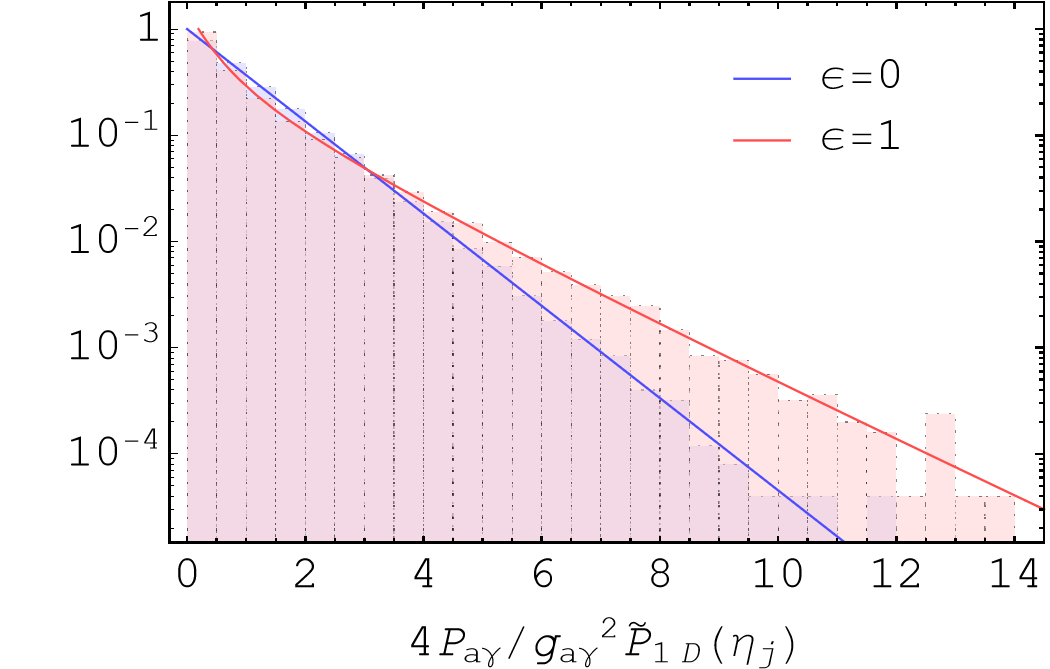}		\caption{Probability distribution of the conversion ratio for fixed values of $g_{a\gamma}$ and $\eta_{j}$ in the case of a Gaussian field (blue line) and the non-Gaussian model (red line). The analytical predictions of Eqs.~\eqref{eq:probdistr0} and \eqref{eq:probdistr} are in excellent agreement with the numerical simulation for $5\times10^{4}$ realisations, $\eta_{j}=0.117~\kpc^{-1}$, and a magnetic field extended over $100$~kpc.}
		\label{fig:probd}
		\end{figure}
Given the PDF $f_{\tilde{B}_{j}}$, we now calculate the probability distribution for the conversion ratio, $P_{a\gamma}$, by imposing the constraint that the probability is proportional to $|\tilde{B}_{j}|^{2}$
\begin{equation}
    \begin{split}
     f_{P_{a\gamma}(\eta_{j})}(p)&=\int db_{1}db_{2}\,f_{\tilde{B}_{j}}(b_{1},b_{2})\delta(b_{1}^{2}+b_{2}^{2}-p)=\\
     &=\int_{-\sqrt{p}}^{\sqrt{p}} db_{2}\,\left[f_{\tilde{B}_{j}}\left(\sqrt{p-b_{2}^{2}},b_{2}\right)+f_{\tilde{B}_{j}}\left(-\sqrt{p-b_{2}^{2}},b_{2}\right)\right]\frac{1}{2\sqrt{p-b_{2}^{2}}}\,,
     \label{eq:probdistr}
    \end{split}
\end{equation}
where $p=4P_{a\gamma}/g_{a\gamma}^{2}\tilde{P}_{\rm 1D}(\eta_{j})$ and $b_1, b_2$ respectively denote the real and imaginary parts of $\tilde{B}_{j}$. The resulting PDFs for the non-Gaussian case with $\epsilon=1$ and the Gaussian case with $\epsilon=0$ are shown in Fig.~\ref{fig:probd}. The non-Gaussian magnetic field produces a characteristic heavy tail, with increased support for high conversion ratios. The heavy tail is reflected in the higher moments of the distribution of $P_{a\gamma}(\eta_j)$:
\begin{equation}
    \begin{split}
        \langle P_{a\gamma}(\eta_j)\rangle&=\left(\frac{\gag^2}{4}  \right)\tilde{P}_{\rm 1D}(\eta_{j})\,,\\
         \langle P_{a\gamma}(\eta_j)^2\rangle&= \left(\frac{\gag^2}{4}  \right)^2
         \tilde{P}_{\rm 1D}^{2}(\eta_{j})\left(3-\frac{4}{(2+\epsilon^{2})^{2}}\right)\,,\\
          \langle P_{a\gamma}(\eta_j)^{3}\rangle&=3\left(\frac{\gag^2}{4}  \right)^3\tilde{P}_{\rm 1D}^{3}(\eta_{j})\left(5-\frac{12}{(2+\epsilon^{2})^{2}}\right)\,,\\
           \langle P_{a\gamma}(\eta_j)^{4}\rangle&=3\left(\frac{\gag^2}{4}  \right)^4 \tilde{P}_{\rm 1D}^{4}(\eta_{j})\left(35+\frac{48}{(2+\epsilon^{2})^{4}}-\frac{120}{(2+\epsilon^{2})^{2}}\right)\,.\\
    \end{split}
\end{equation}
The mean of the conversion ratio is unchanged as the power spectrum of the magnetic field is, by construction, the same for the Gaussian and non-Gaussian magnetic fields. However, the higher-order moment 
all increase for $\epsilon\neq0$ due to non-Gaussianities, which is indicative for the heavy-tailed probability distribution. 
		
\section{Heavy tails for lighter axions}
\label{App:massless}
In the bulk of this paper, we have considered the case of $m_a>\omega_{\rm pl}$ in which numerical simulations can be matched by detailed analytical results. In this appendix, we extend this discussion to the case of arbitrary $\omega_{\rm pl}(z)/m_a$ and show that, also in this case, MHD magnetic fields lead to heavy-tailed probability distributions.

The ALP--photon transition amplitude is, to leading order in perturbation theory, given by~\cite{Raffelt:1985nj, Marsh:2021ajy}
\begin{equation}
   i {\cal A}_{\gamma_{i}\to a} = \frac{g_{a\gamma}}{2} \int_{-\infty}^{\infty} dz\,
    B_i(z\hat z) e^{i\lambda\phi(z)} \,,
    \label{eq:gen}
\end{equation}
where
\begin{equation}
    \phi(z)=\int_{0}^{z} dz'\frac{\omega_{\rm pl}^{2}-m_{a}^{2}}{2}\,,
\end{equation}
and $\lambda=1/\omega$. The function $\phi(z)$ is not, in general, linear in $z$, and is even non-monotonous if there are `resonance points' (i.e.~level-crossings) where $\omega_{\rm pl}-m_a$ changes sign along the trajectory. We note that the amplitude can still be expressed as a sum of Fourier transforms (of $B_i/|\phi'|$) and efficient numerical evaluation is possible through methods like the Fast Fourier Transform, though we will not use these methods here. The new conceptual issue of this general case is that regions where $|\phi'|$ is small are expected to contribute more to the amplitude than similar regions with large $\phi'$. Naively, one might expect that resonance points give large contributions to the amplitude, and that the amplitude would be rather insensitive to the magnetic properties far away from the resonance points. If so,  non-Gaussian peaks of the magnetic field would only contribute significantly if located at a resonance point, which could lead to very different properties of the distribution of the conversion ratio.  We now show that this intuition is not correct: resonant contributions are typically small compared to the cumulative, non-resonant contributions, and the heavy tails of the non-Gaussian magnetic field persist also in this general case.

From our MHD runs, we first extract the plasma frequency as a function of position. The average plasma frequency value for the two Runs is $\bar \omega_{\rm pl}=6.59\times10^{-12}$~eV, and fluctuations around the mean are small, indeed much smaller than the expected range of the plasma frequency in a galaxy cluster. We therefore re-scale the fluctuations to an amplitude of $\sim3 \bar \omega_{\rm pl}$ around the mean value and choose to consider ALP masses that have one or more resonance points within the simulation box. We then calculate the conversion ratio as described in the main text of this paper. 

\begin{figure*}[t!]
		\vspace{0.cm}
		\includegraphics[width=0.45\linewidth]{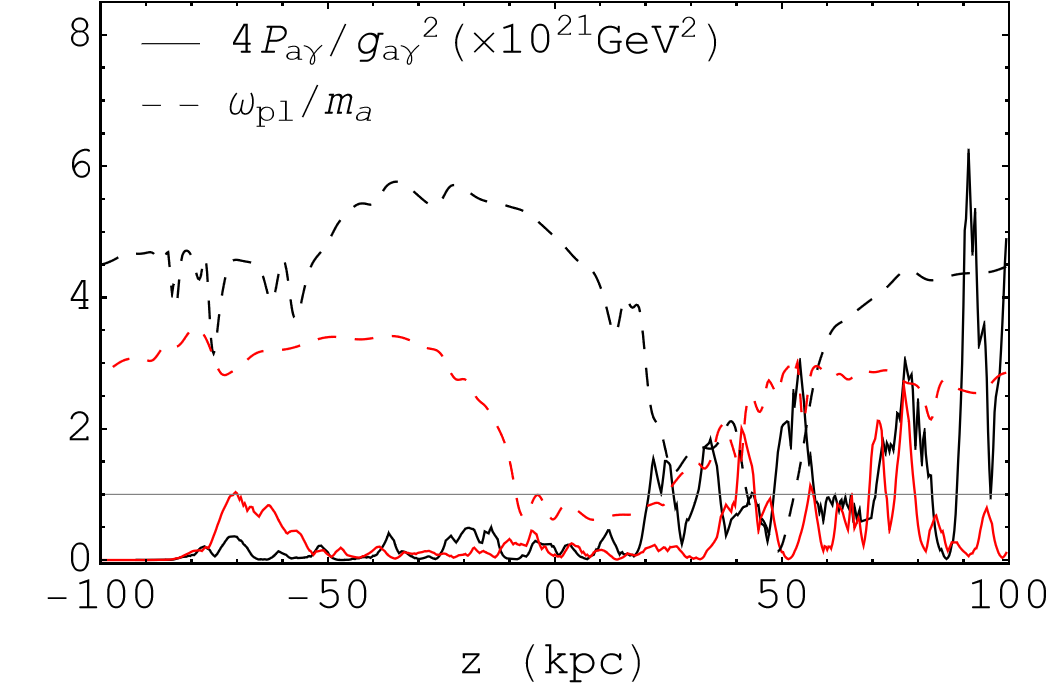}
\caption{Conversion ratio (solid lines) as function of the distance along the $z$ axis in Run $\mathcal{S}$ for two trajectories represented by the two colours. The ALP mass is $m_{a}=6.59\times10^{-12}$~eV, the energy $\omega=10$~keV and the dashed lines show $\omega_{\rm pl}(z)/m_{a}$. The grey line indicates the resonance condition, $m_{a}=\omega_{\rm pl}$.
}\label{fig:full}
\end{figure*}

Figure~\ref{fig:full} shows the conversion ratio as a function of $z$ along two trajectories within Run $\mathcal{S}$. Each trajectory has two resonance points where $\omega_{\rm pl}(z)/m_{a}=1$. Clearly, around the resonance points, there is no significant increase in the conversion ratio which, as in the massive case, is determined by the properties of the magnetic field along the entire trajectory. This result is consistent with the general discussion in \cite{Marsh:2021ajy}.

\begin{figure*}[t!]
		\vspace{0.cm}
		\includegraphics[width=0.45\linewidth]{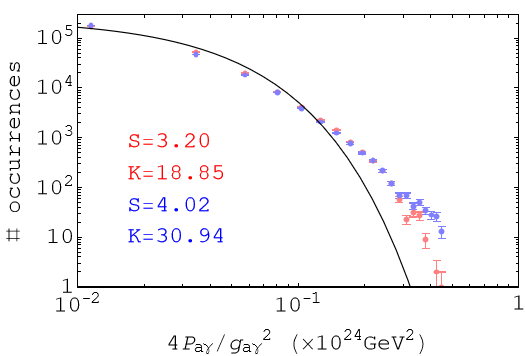}
		\includegraphics[width=0.45\linewidth]{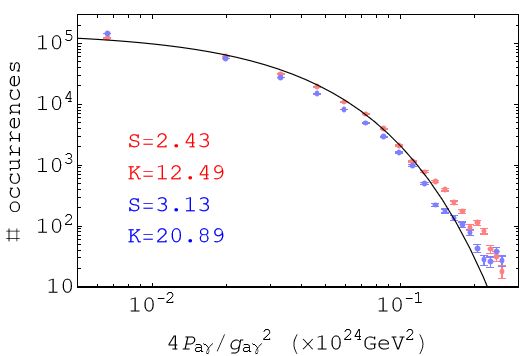}	
\caption{
Histogram of the conversion ratio for $m_a = \bar \omega_{\rm pl} = 6.59\times10^{-12}$~eV extracted from Runs $\mathcal{K}$ and $\mathcal{S}$ for an initial photon flux that is linearly polarised along the $x$ (blue) and $y$ (red) directions together with the analytical GRF prediction for the case of a \emph{constant} plasma frequency (black line). Here $\omega=10\,{\rm keV}$. The error bars are obtained by assuming Poissonian fluctuations.}\label{fig:probdMHDmassless}
\end{figure*}

Figure~\ref{fig:probdMHDmassless} shows the numerical  distribution of the conversion ratio, $P_{a\gamma}$, in the MHD runs with $m_a=\bar \omega_{\rm pl}$. The solid lines indicate the analytical predictions for the same ALP mass but considering the   massive case (discussed in the main text) where the plasma frequency is negligible. The comparison of the analytical result (solid line) and the numerical simulations (histograms) is therefore not rigorous, but the appearance of heavy tails at large conversion ratios is apparent in both figures. We checked that similar results are obtained in case of very light ALPs, $m_{a}\ll\omega_{\rm pl}$. This indicates that the conclusions of this work hold over the entire ALP parameter space.

\section{Origin of the heavy tails}
\label{App:whyfat}
\begin{figure*}[ht!]
		\vspace{0.cm}
		\includegraphics[width=0.45\linewidth]{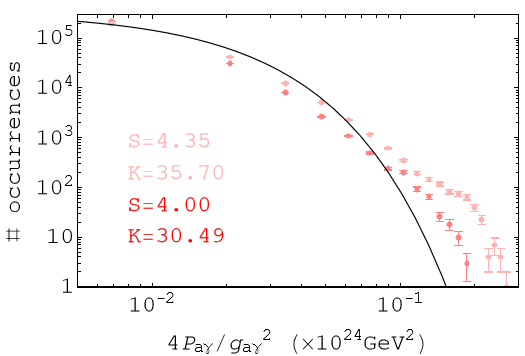}
		\includegraphics[width=0.45\linewidth]{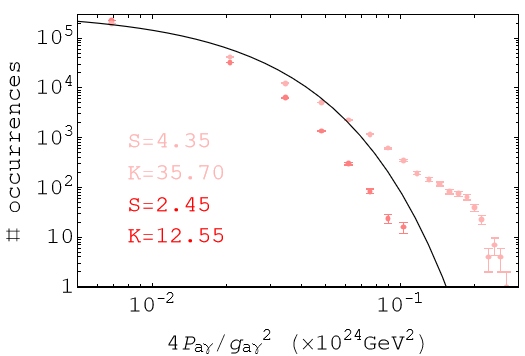}
		\caption{		Histogram of the conversion ratio extracted from Run $\mathcal{K}$ of the MHD simulation (light red) for a linearly polarised initial photon flux together with the analytical GRF prediction (black line). Here $\eta_{a}=0.07~\kpc^{-1}$.
		The red histogram obtained by: masking regions where $L_{B}>5$~kpc (left panel); masking regions where $|B_{\perp}|>7~\muG$ (right panel). The error bars are obtained by assuming Poissonian fluctuations.
		}
		\label{fig:HISTCUT}
		\end{figure*}

In the following we discuss what are the reasons behind the heavy-tails of distributions of axion-photon conversions.
We identify two distinct non-Gaussian features of MHD magnetic fields that might be at the origin of the heavy tails of $f_{P(\eta_a)}$: large coherent structures, and larger amplitude peaks (spikes) in the modulus of the magnetic field.
Figure~\ref{fig:HISTCUT} shows the empirical distributions $f_{P(\eta)}$ obtained after masking regions with large coherent structures. Specifically, we have defined the magnetic length scale as
\begin{equation}
    L_{B}=\left|\frac{dz}{d\ln |\bB|}\right|\,,
\end{equation}
where the derivative is taken along the direction of propagation. We have then removed the contribution from regions where $L_B>5$~kpc, resulting in the histograms in the left panel of Fig.~\ref{fig:HISTCUT}. The effect is a strong convergence of the skewness and kurtosis towards the GRF prediction (again, $S=2$ and $K=9$).
To investigate the role of spikes in $\bB$, we mask regions where the transverse components' magnetic field is $|B_{\perp}|>7\muG\sim 2 B_{\rm rms}$ and the present the result in the right panel of Fig.~\ref{fig:HISTCUT}. There is a drastic reduction in the skewness and kurtosis, and the predictions closely approximate those of the GRF. The cuts are chosen in such a way that photons traverse a sufficiently large region with non-vanishing magnetic field.
Clearly, the independence of $P_{a\gamma}$ on the phases of $\tilde B$ suffices to make the expectation value of the conversion ratio independent of MHD structure but does not guarantee that higher-order correlation functions of $P_{a\gamma}$ are Gaussian, or that the full PDF follows the exponential form.


\section{Unpolarized beams and helicity}
\label{App:helicity}
In this appendix, we discuss the observationally interesting case of unpolarised photon fluxes. The relevant conversion ratio is given by $P_{\gamma a}(\eta_a)=\frac{1}{2} \left(P_{\gamma_x  a}(\eta_a)+P_{\gamma_y  a}(\eta_a) \right)$, so that
\begin{equation}
     P_{a\gamma}(\eta_{a})= \frac{1}{2} \frac{g_{a\gamma}^{2}}{4} \int\frac{d^{2}\bk_{\perp}}{(2\pi)^{2}}\frac{d^{2}\bk_{\perp}'}{(2\pi)^{2}} \left[ B_{x}(\eta_{a}\bn+\bk_{\perp})B_{x}^{*}(\eta_{a}\bn+\bk_{\perp}') + B_{y}(\eta_{a}\bn+\bk_{\perp})B_{y}^{*}(\eta_{a}\bn+\bk_{\perp}') \right] \, .
\end{equation}
Clearly, the expectation value of the unpolarised conversion ratio is immediately determined by the corresponding polarised conversion ratios. However, higher moments of the unpolarised conversion ratio will involve cross-correlators of $\hat B_x$ and $\hat B_y$, which are non-vanishing in the presence of helicity, cf.\ Eq.~\eqref{eq:2point}. The first two moments of the conversion ratio are explicitly given by
\begin{equation}
    \begin{split}
        \langle  P_{a\gamma}(\eta_{a})\rangle&=\frac{g_{a\gamma}^{2}}{4}\tilde{P}_{\rm 1D}(\eta_{a})\,,\\
        \langle  P_{a\gamma}^{2}(\eta_{a})\rangle&= \frac{3}{2}\left(\frac{g_{a\gamma}}{2}\right)^{4}\tilde{P}_{\rm 1D}(\eta_{a})^{2}+\frac{1}{2}\left(\frac{g_{a\gamma}}{2}\right)^{4}f_H^2(\eta_{a}) \,,
    \end{split}
    \label{eq:2point}
\end{equation}
where we defined
\begin{equation}
\begin{split}
    f_H^2(\eta_{a}) &=\left[\frac{L}{2\pi} \int \frac{dk_{\perp}\,k_{\perp}}{(2\pi)^{3}} \frac{\eta_{a}}{\sqrt{\eta_{a}^{2}+k_{\perp}^{2}}}H(\eta_{a},k_{\perp}) \right]^{2}\,.
\end{split}
\end{equation}
More generally, for GRFs the helicity always enters the correlation functions through an even power of $f_H$, and hence gives non-negative contributions to higher moments, and can contribute to  heavy tails of the probability distribution of $P_{a\gamma}$ (cf.~the non-Gaussian example of appendix \ref{App:nonG}. For non-Gaussian fields, the situation is model dependent as there are additional contributions involving powers of $f_H$ and correlators of an odd number of magnetic fields. However, as we will now argue, in practice helicity tends to be negligible for ALP--photon conversion also for MHD fields.

		\begin{figure}[t!]
		\vspace{0.cm}
\includegraphics[width=0.45\linewidth]{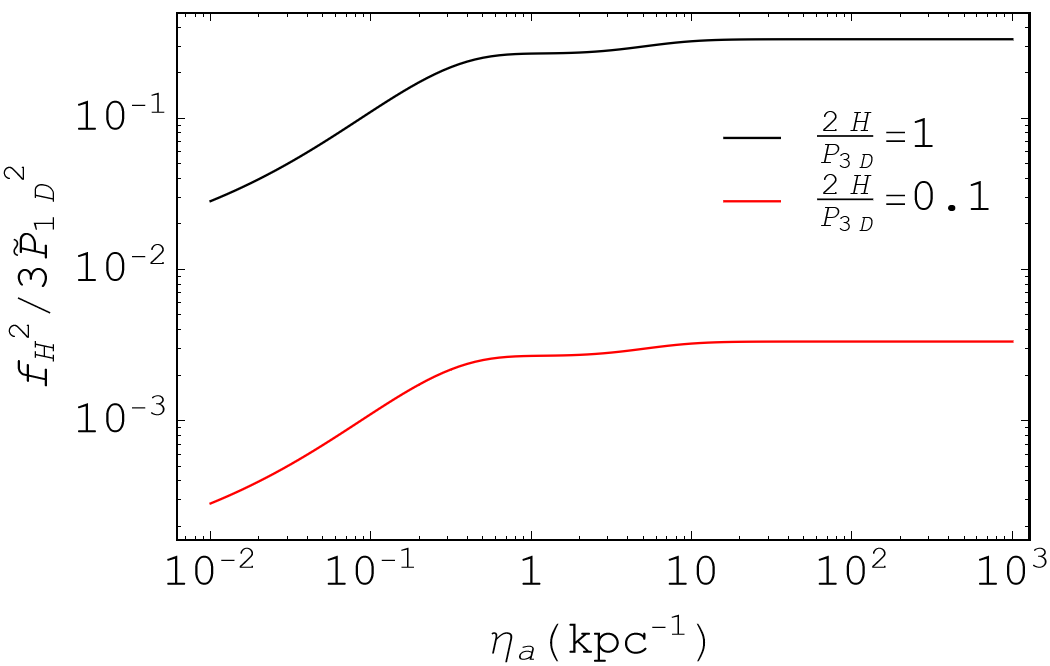}
		\caption{The ratio of $f_H^2$ (contribution of magnetic helicity spectrum to the ALP to photon conversion ratio square) to $3\tilde{P}_{\rm 1D}^2$ (contribution from transverse magnetic power spectrum) as a function of $\eta_{a}$ for different values of helicity fraction.}
		\label{helical_contribution}
		\end{figure}
In general, the magnetic fields generated by small-scale dynamo with non-helical forcing are non-helical. To provide a quantitative comparison for the role of helicity of the magnetic field in $P_{a\gamma}$, we consider a scenario where the helicity spectrum ($H$) is proportional to the power spectrum ($M_{N}=P_{\rm 3D}/2$) and the ratio $2H/P_{\rm 3D}$ is either 1 i.e.~the fully helical magnetic field case or 0.1. Figure~\ref{helical_contribution} shows the ratio $f_{H}^{2}/3P_{\rm 1D}^{2}$ for these two cases. To estimate this ratio we considered a 3D power spectrum given by~\cite{Axel2022}
\begin{equation}
  P_{\rm 3D}(\eta_{a},k_{\perp})=\frac{1}{1+\left(\frac{k}{k_{p}}\right)^{11/3}}\,, \quad k=\sqrt{\eta_{a}^{2}+k_{\perp}^{2}}\,,
 \end{equation}
where $k_p=(2\pi/10) \text{kpc}^{-1}$ is the wavenumber corresponding to the peak of the spectrum. From Fig.~\ref{helical_contribution}, it is concluded that the impact of the helicity on the photon to ALP conversion ratio is negligible unless the magnetic field is fully helical and $\eta_{a}>k_p$. In applications it means that it is more important for low ALP energy, where typically the ALP--photon conversions are suppressed. 

\section{Magnetohydrodynamic simulations}
\label{App:MHD}
In this appendix, we review details of the properties of the MHD simulations.
Magnetic fields in galaxy clusters are thought to be generated through gas motions by a dynamo process. The gas motions are turbulent, but their driving is not well understood. Magneto-thermal instabilities have been studied in this context~\cite{Scheko+05, Parrish+12}, and even the driving by turbulent wakes of individual galaxies has been discussed~\cite{Ruzmaikin+89,Subramanian+06}. Turbulence could also be driven by sporadic cluster mergers~\cite{Xu+09,Vazza+12}, but the turbulence would be slowly decaying between such events.

The dynamo process itself is a generic phenomenon that is mainly
characterized by a large magnetic Prandtl number~\citep{BS05}.
This means that the magnetic diffusivity $\eta$ is much smaller than
the kinematic viscosity $\nu$.
This is numerically difficult to handle.
In addition, the magnetic Reynolds number is huge (of the order of
$10^{30}$).
This means that the turbulent magnetic cascade extends down to very
small length scales.
As a compromise, since the energy contained in the smallest length
scales becomes very weak, we consider here turbulence at a moderate
magnetic Reynolds number that is as large as possible for a given
numerical resolution.
The turbulence is assumed to be driven through some external volume
forcing as is commonly done in numerical simulations of homogeneous
turbulence.
We also ignore the density stratification and just consider an isothermal
gas with constant sound speed.

As advertised above, we consider a dynamo-generated magnetic field
$\BB$ driven by forced turbulence.
We solve the induction equation for the magnetic vector potential $\AAA$
with $\BB=\nab\times\AAA$ in the Weyl gauge, i.e.
\begin{equation}
\frac{\partial\AAA}{\partial t}=\uu\times\BB-\eta\JJ\,.
\label{eq:dAdt}
\end{equation}
Here, we have adopted Gaussian Heaviside units where $\JJ=\nab\times\BB$
is the current density, and $\uu$
is the velocity which obeys the momentum equation,
\begin{equation}
\frac{\DD\uu}{\DD t}=-\cs^2\nab\ln\varrho+\ff+\frac{1}{\varrho}
\left[\JJ\times\BB+\nab\cdot(2\varrho\nu\SSSS)\right]\,,
\label{eq:dudt}
\end{equation}
where $\DD/\DD t=\partial/\partial t+\uu\cdot\nab$ is the adjective time derivative, ${\sf S}_{ij}=(\partial_j u_i+\partial_i u_j)/2 -\delta_{ij}\nab\cdot\uu/3$ are the components of the traceless rate-of-strain tensor $\SSSS$, $\ff$ is the forcing function, and $\varrho$ is the density, which obeys the continuity equation, written here as
\begin{equation}
\frac{\DD\ln\varrho}{\DD t}=-\nab\cdot\uu\,.
\label{eq:dlnrhodt}
\end{equation}
We solve these equations in a periodic domain of size $L^3$ with $512^3$
mesh points using the {\sc Pencil Code}~\cite{JOSS}.
The forcing function consists of random sinusoidal waves that are
$\delta$-correlated in time, i.e.~the forcing function changes at each
time step.
The wave vectors are taken from a shell of finite thickness and radius
$\kf$, which we chose to be close to the smallest wave number of the
computational domain $k_1\equiv2\pi/L$.
The strength of the forcing function is arranged such that the turbulent
Mach number $\Ma=\urms/\cs$, where $\urms$ is the rms velocity,
is around unity or less.
We define the kinetic and magnetic Reynolds numbers as
$\Rey=\urms/\nu\kf$ and $\Rm=\urms/\eta\kf$, respectively,
and the Lundquist number as $\Lu=\vA^{\rm rms}/\eta\kf$,
where $\vA^{\rm rms}$ is the rms value of the magnetic field
in Alfv\'en velocity units, $\vvA=\BB/\sqrt{\varrho}$.
Assuming Kolmogorov scaling, our nominal viscous and resistive cutoff
wavenumbers are $k_\nu=\sqrt{\omega_{\rm rms}/\nu}$ and
$k_\eta=k_{\nu}(\nu/\eta)^{-1/2}$,
respectively, where
$\omega_{\rm rms}$ and $\omega_{\rm A}^{\rm rms}$ are the rms values of
the vorticities $\nab\times\uu$ and $\nab\times\vvA$, respectively.
The basic parameters of the models are listed in Table~\ref{Tsum}.
We also quote the average of the kurtosis of the three components
of the magnetic field, $\kurt B_i=\bra{B_i^4}/\bra{B_i^2}^2$
for $i=1,2,3$.
For a Gaussian distributed field, $\bra{\kurt B_i}=0$.

We initialise the simulations with a weak seed magnetic field. After
about 50 turnover times ($\urms\kf t=10$), the magnetic field begins
to grow exponentially.
During this phase, the magnetic field is highly non-Gaussian, but the
field strength is still weak.
To assess the consequences of such a highly non-Gaussian field, we
consider a scaled version of this magnetic field, referred to as Run $\mathcal{K}$,
because the dynamo is kinematic, i.e.~unaffected by magnetic feedback.

When the magnetic energy density reaches values comparable to the kinetic
energy density, the Lorentz force $\JJ\times\BB$ begins to affect the
turbulence and leads to a saturation of the dynamo (Run $\mathcal{S}$).
The magnetic field is then still non-Gaussian, but the kurtosis is smaller
than during the kinematic stage.
The density also becomes more strongly affected by the magnetic field.
A cross-section of the magnetic field for Run $\mathcal{S}$ is shown in the left
panel of Fig.~\ref{fig:Hfig} and also for the case where magnetic fields
are considered to be GRF (middle panel) with the same power spectrum as
in Run $\mathcal{S}$.
The red regions in these slices highlight locations with
$|B|>3\Brms$. There are more such regions in Run $\mathcal{S}$ compared
to the GRF case, as is evident from the figure.
We explore the role of such regions in the photon to ALP conversion
probability in the later section.

\begin{table}[t!]\caption{Parameters of each magnetic field model used.
}\vspace{12pt}\centerline{\begin{tabular}{cccccccc}
Run & $\Ma$ & $\Rey$ & $\Rm$ & $\Lu$ &  $k_\nu/k_1$ & $k_\eta/k_1$ & $\bra{\kurt B_i}$ \\
\hline
${\mathcal K}$ & 0.12 & 75 & 1500  & 0 & 20 & 90 & 17.1 \\
${\mathcal S}$ & 0.08 & 55 & 1100 & 910  & 17 & 72 &  5.8 \\
\label{Tsum}\end{tabular}}\end{table}

\section{Single-sightline statistics}
\label{app:statist}
Most observational probes are directly sensitive only to the magnetic field along a single sightline from the source to the detector, but the heavy-tailed distribution we have uncovered is evident in the statistical distribution of an ensemble of sightlines, probed at fixed energy. It is natural to expect that the large conversion probabilities of the heavy tails correspond to large-amplitude oscillations in energy-dependent photon spectra, and in this Appendix, we demonstrate how this works for a mock X-ray observation. 
		\begin{figure}[t!]
		\vspace{0.cm}
\includegraphics[width=0.45\linewidth]{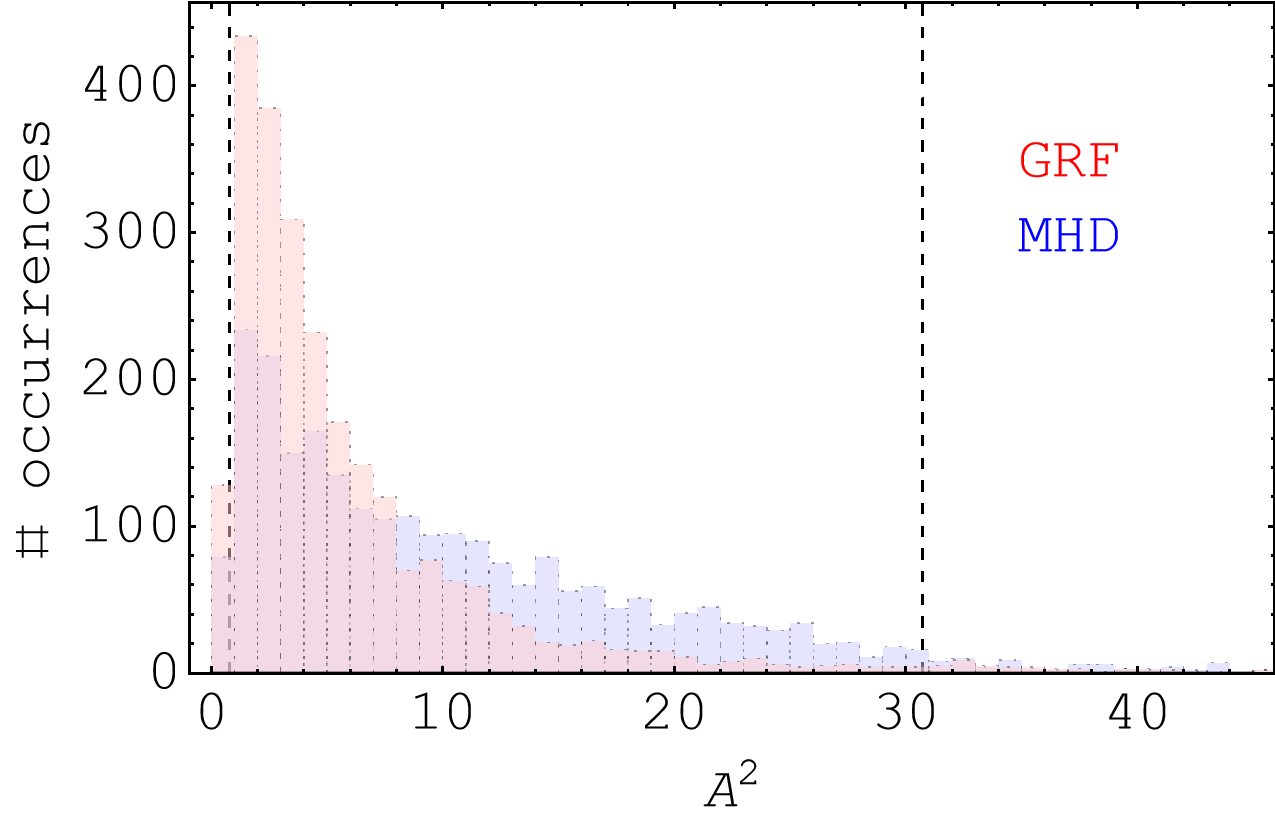}
\includegraphics[width=0.45\linewidth]{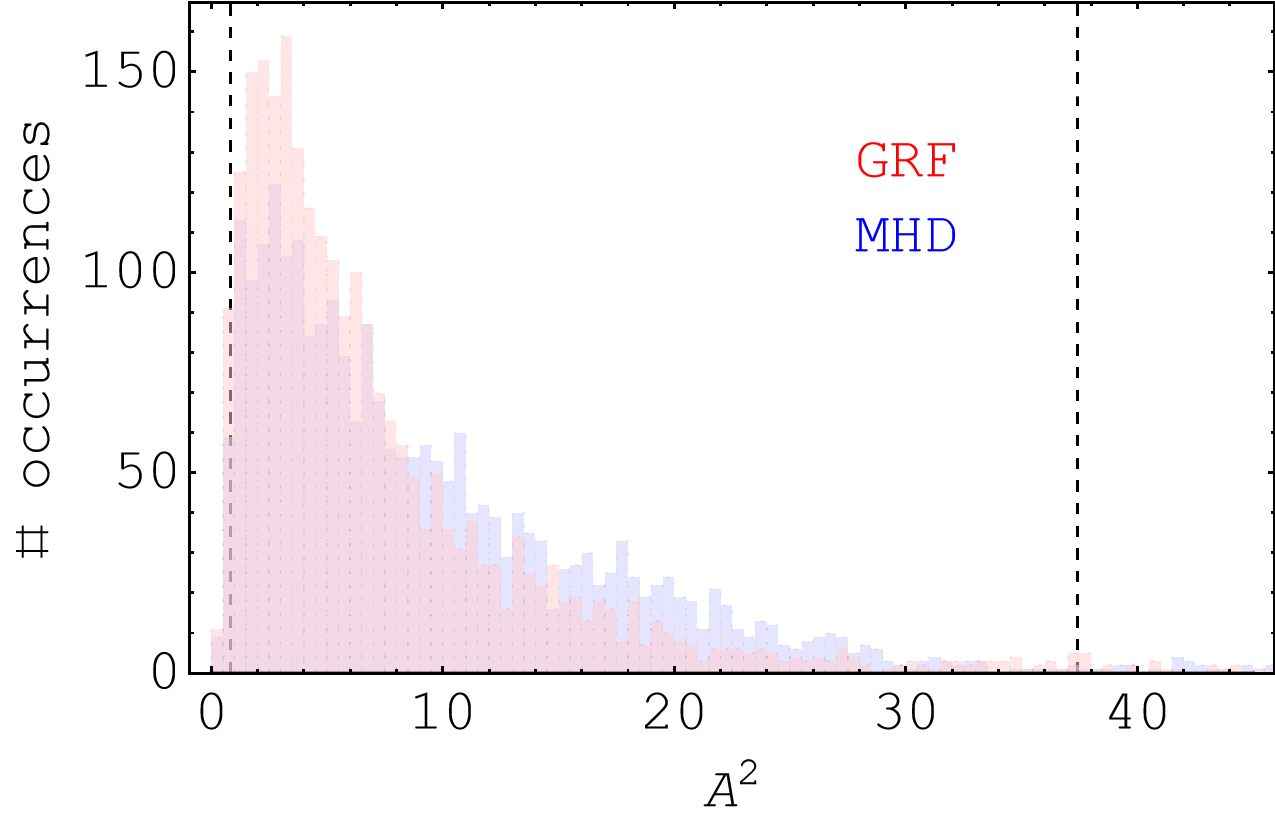}
		\caption{Comparison of the distribution for the Anderson-Darling variable $A^{2}$ for GRF models (red histograms) with the same power spectrum of run $\mathcal{K}$ (blue histogram, left panel) and run $\mathcal{S}$ (blue histogram, right panel) models. The vertical black dashed lines denote the $95\%$~CL interval for the test variable.
  Here we used $\Delta E=4.7$~eV, $m_{a}=10^{-12}$~eV and $g_{a\gamma}=5\times10^{-13}~\GeV^{-1}$.}
		\label{fig:A2}
		\end{figure}
We consider hypothetical X-ray observations of a bright, power-law source in a galaxy cluster, observed by instruments with energy resolutions comparable to current and future experiments. We proceed as follows:
\begin{itemize}
    \item We consider a source with intrinsic power-law spectrum with index $-2$, and modulate these by the energy-dependent conversion ratios calculated in the MHD and GRF model.  The observed photon flux is tabulated as a function of the energy for a set of energy-spacings, where a finer binning roughly corresponds to an X-ray experiment with higher energy resolution. The final results are insensitive to the value of the spectral index.
    \item The resulting spectrum is fitted with a power law, and the residuals are extracted.
    \item For the GRF model, we determine the distribution of the residuals by scanning 2,500 realisations.
    \item For the residuals from each MHD realisation, we test the null hypothesis that the residuals come from the GRF distribution. 
\end{itemize}
The resulting cumulative distribution function (CDF) of the residuals for a GRF is shown in Fig.~\ref{fig:CDFs} (dashed lines) and compared with single realisations of each of the two MHD models considered (solid lines). 
Intuitively, one may expect more negative residuals for the MHD case compared to the GRF model since the heavy tails give larger conversion ratios and deeper `dips' in the observed spectrum. This effect is indeed visible in Fig.~\ref{fig:CDFs}, where the empirical CDFs of the MHD realisations grow faster than the GRF CDF. 

We use the Anderson-Darling test to determine the significance of the heavy tails for single-sightline spectra.  
%
%
The Anderson-Darling test is a non-parametric statistical test that can be used to compare a sample with a theoretical model, to determine whether the sample is generated by the theoretical underlying distribution. This test is more sensitive to differences in the distributions' tails than other distributional equality tests, such as the Kolmogorov-Smirnov test. This makes it particularly useful for detecting differences in extreme values. Its non-parametric nature and sensitivity to differences in the tails of the distributions make it a valuable tool in this context. We determine the distribution of the test statistic~\cite{doi:10.1080/01621459.1987.10478517}
\begin{equation}
A^{2}=-n-\frac{1}{n} \sum_{i=1}^{n}\left[ \frac{2i - 1}{2n} \ln(F(Y_{i})) + \ln(1 - F(Y_{n-i+1})) \right]\,,
\end{equation}
where $n$ is the sample size and $Y$ is an ordered list of residuals. This step allows us to determine the 95\% confidence level for the variable $A^{2}$. These results are shown in Tab.~\ref{tab:interv} for different values of the energy resolution and the two magnetic field models considered. For illustrative purpose, we show the distribution of the statistical test variable $A^{2}$ in Fig.~\ref{fig:A2}, comparing the GRF case (red histograms) with run $\mathcal{K}$ (blue histogram, left panel) and run $\mathcal{S}$ (blue histogram, right panel). Slightly larger values of $A^{2}$ are preferred by the MHD models compared to the GRF case.

Note that we consider $N_{\rm div}$ bins, determining an equally spaced grid in terms of $\Log_{10}\omega$, in the $0.1-10$~keV range in which we calculate the simulated photon signal. The energy resolution $\Delta E$ (first column of Tab.~\ref{tab:interv}) is the smallest energy interval that we simulate.
		\begin{figure}[t!]
		\vspace{0.cm}
\includegraphics[width=0.45\linewidth]{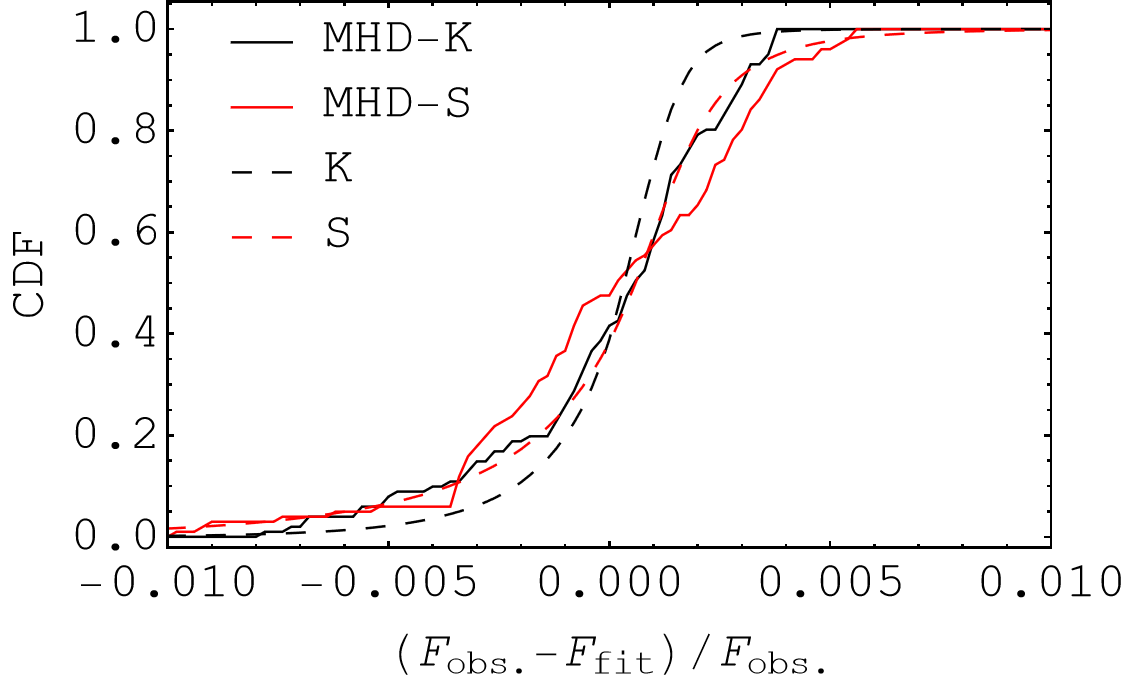}
		\caption{Comparison of the theoretical CDFs (dashed lines) extracted from GRF models with the same power spectrum of run $\mathcal{K}$ (black lines) and run $\mathcal{S}$ (red lines) models. This prediction is compared with two CDFs extracted from the two MHD models considered (solid lines). Here we used $\Delta E=4.7$~eV, $m_{a}=10^{-12}$~eV and $g_{a\gamma}=5\times10^{-13}~\GeV^{-1}$.}
		\label{fig:CDFs}
		\end{figure}

\begin{table}[t!]
    \centering
    \begin{tabular}{|c|c|c|}
    \hline
      $\Delta E~(\eV)$   & $\mathcal{K}$& $\mathcal{S}$\\
      \hline
      58.5& 0.23-3.93&0.24-4.59\\
      9.6 & 0.45-15.75 & 0.48-19.03 \\
      4.7& 0.78-30.71&0.84-37.43\\
        \hline
    \end{tabular}
    \caption{Critical intervals for the stochastic variable $A^{2}$ for different energy resolutions.  The first column refers to the energy resolution that we use to simulate the detected signal. The 95\% of the GRF realizations give a value of $A^{2}$ in the range in the second and third columns of the table, referring to the two different MHD models. Here the ALP parameters are $m_{a}=10^{-12}$~eV and $g_{a\gamma}=5\times10^{-13}~\GeV^{-1}$.}
    \label{tab:interv}
\end{table}

The critical values in Tab.~\ref{tab:interv} are used to determine the similarity between a MHD realization and the GRF prediction. If the test statistic is outside the interval determined by the critical values, the null hypothesis that the the MHD residuals come from the same distribution of the GRF ones, is rejected. This comparison is performed over 2500 lines-of-sight for the MHD models. These lines-of-sight are uniformly distributed in the innermost $190$~kpc of the simulation box and connect two opposite faces of the box.

Finally, we count the number of times the hypothesis that the residuals are extracted from the same distribution of the GRF is rejected at the $95\%$ confidence level. Therefore, we can estimate the probability that a $P_{a\gamma}$ vs $\omega$ realization of an MHD run is significantly different from the GRF case. These results are shown in Tab.~\ref{tab:tabconcl}.  This analysis allows us to conclude that, even with high energy resolution, only in the $5-8\%$ of the cases the observed line-of-sight differs significantly from the predictions of a GRF model. This suggests that GRF models of turbulent cluster magnetic fields suffice to reliably determine the predictions for X-ray searches for ALPs for observables probing a single sightline.
\begin{table}[t!]
    \centering
    \begin{tabular}{|c|c|c|}
    \hline
      $\Delta E~(\eV)$   & $\mathcal{K}$ (\%)& $\mathcal{S}$ (\%)\\
      \hline
      58.5&8.3&5.7\\
      9.6 &7.9 &4.8 \\
      4.7&8.0 &4.9\\
        \hline
    \end{tabular}
    \caption{Probability that a line-of-sight of an MHD model is significantly different (as defined in the text) from the GRF case. The first column refers to the energy resolution that we use to simulate the detected signal, the second and third ones to the probability for the two MHD models considered. Here the ALP parameters are $m_{a}=10^{-12}$~eV and $g_{a\gamma}=5\times10^{-13}~\GeV^{-1}$.}
    \label{tab:tabconcl}
\end{table}

\bibliographystyle{bibi}
\bibliography{biblio.bib}

\end{document}